\documentclass[aps,prd,twocolumn,groupedaddress,floatfix,
longbibliography,superscriptaddress]{revtex4-2}
\usepackage{amsmath,amssymb,amsfonts,mathrsfs,bm,bbold,braket,slashed,cancel}
\usepackage{graphicx}
\usepackage{epsfig}
\usepackage{float}
\usepackage{subfigure}
\usepackage{caption}
\usepackage[labelformat=simple]{subcaption}
\usepackage{multirow}
\usepackage{comment}
\usepackage[usenames,dvipsnames,svgnames]{xcolor}
\usepackage{hyperref}
\usepackage{etoolbox}
\usepackage{orcidlink}
\hypersetup{
    colorlinks=true,
    linkcolor=Blue,
    citecolor=Blue,
    urlcolor=Blue
}
\usepackage{accents}
\usepackage[normalem]{ulem}
\usepackage{simpler-wick}
\usepackage{tikz}
\usetikzlibrary{tikzmark, positioning, fit, shapes.misc}
\tikzstyle{line}=[draw]
\usepackage[compat=1.1.0]{tikz-feynman}
\renewcommand{\Im}{\operatorname{Im}}
\renewcommand{\Re}{\operatorname{Re}}
\newcommand{\bea}{\begin{eqnarray}}
\newcommand{\eea}{\end{eqnarray}}
\newcommand{\be}{\begin{equation}}
\newcommand{\ee}{\end{equation}}

\renewcommand\vec{\bm}
\renewcommand\Re{\text{Re}}
\renewcommand\Im{\text{Im}}

\begin{document}
\setlength{\belowdisplayskip}{10pt}


\title{Electric field fluctuations and renormalization group flows in a self-interacting scalar field theory}

\author{Melanie Martínez Villarreal}
 \email{mimartinez1@uc.cl}
 \affiliation{Facultad de Física, Pontificia Universidad Católica de Chile, Vicuña Mackenna 4860, Santiago, Chile}
\author{Enrique Mu\~noz~\orcidlink{0000-0003-4457-0817}}
 \email{corresponding author: ejmunozt@uc.cl}
 \affiliation{Facultad de Física, Pontificia Universidad Católica de Chile, Vicuña Mackenna 4860, Santiago, Chile}
 \affiliation{Center for Nanotechnology and Advanced Materials CIEN-UC, Avenida Vicuña Mackenna 4860, Santiago, Chile}
 \author{Marcelo Loewe}
 \email{marcelo.loewe@uss.cl}
 \affiliation{Facultad de Ingeniería, Universidad San Sebastián, Santiago, Chile}
 \affiliation{Centre for Theoretical and Mathematical Physics, and Department of Physics, University of Cape Town, Rondebosch 7700, South Africa}

\date{\today}

\begin{abstract}
We consider a self interacting charged scalar field represented by the complex $\lambda\phi^4$ model, embedded in a background electric field exhibiting classical stochastic fluctuations. We studied the effects of the classical stochastic noise on the physical parameters of the scalar field theory, for both weak and ultra strong electric field regimes. The stochastic background electric field is included in the Schwinger propagator through the covariant derivative, and the generating functional of the theory is found by means of the replica trick in order to compute the statistical average over electric fluctuations. As a result of the averaging process, an effective interaction between charged currents emerges, with a coupling constant proportional to the magnitude of the auto-correlation function of the electric field fluctuations. We obtained the dressed propagators, the interaction vertices, and the renormalization group equations of the theory, along with the corresponding streamplots in the manifold of interaction couplings. 
\end{abstract}

\maketitle


\section{\label{sec:level1}Introduction}
Complex scalar fields are ubiquitous in effective quantum field theories, ranging from the analysis of critical phenomena in condensed matter physics and statistical mechanics to the description of charged boson mesons such as pions~\cite{Baker_1960}. In this second context, phenomenological scenarios in heavy-ion collisions~\cite{Busza_2018,Iancu_2012} and astrophysics involve the presence of background electromagnetic fields~\cite{Deng_2015} that interact with these charged bosons. In non-central relativistic heavy-ion collisions, electromagnetic fields arise due to the spectator nucleons outside the overlapping region. Particularly, in the case of asymmetric collisions such as Cu+Au, a stronger dipole-like electric field emerges due to the asymmetric distribution of the nuclear charge of the spectators~\cite{cheng_asymmetric}. Although this electric field is present on time-scales of at most $\sim 1$ fm/c~\cite{toneev-directed-flow,hirono-electric-cond}, its magnitude can reach average values up to $\left<qE\right> \sim 10m_\pi^2$ for Au+Au collisions at RHIC \cite{alam-efield}. In those collisions, the production of pions has been predicted~\cite{Gao_PRD_2021} and detected~\cite{Adamczewski_EPJA_2020}. The electromagnetic fields generated at the collision region are expected to interact with the quark-gluon plasma, and influence the dynamics of the thousands of particles~\cite{Kumar_PRD_2025} produced after hadronization.\\

Historically, most analytical approaches on this matter are based on the seminal work of Schwinger, which provides an integral representation of scalar and fermion propagators in terms of the so-called Schwinger proper time~\cite{Schwinger_1951}, in the presence of a constant and uniform electromagnetic field. More recently, some of us proposed a formalism to study the effects of classical stochastic fluctuations over an otherwise constant and uniform magnetic field background on the fermion propagator~\cite{Castano_Munoz1_2023}. Interestingly, the physical picture revealed that under a classical background magnetic field with stochastic fluctuations, the fermions propagate in an effective dispersive medium with an anisotropic refraction index for the group velocity~\cite{Castano_Munoz1_2023}. Moreover, the physical consequences of this theory were explored, particularly concerning the effects on the Fermion mass~\cite{Castano_Munoz1_2024}, the generation of a screening effective mass on the photon propagator~\cite{Castano_Munoz2_2024}, and more recently the production of photons~\cite{Castano_Munoz_2025} under such a fluctuating magnetic medium.

In the present paper, we shall apply a similar analysis for the propagator of a charged scalar field immersed in a classical electric field background with stochastic fluctuations. As a toy model for charged pions in this fluctuating medium, we consider a self-interacting $\lambda \phi^4$ theory, that has been used in the past to analyze the main features of $\pi-\pi$ scattering processes~\cite{Baker_1960}. Moreover, the coupling of this model to a constant background electric field, in the effective potential formalism for chiral symmetry breaking, has been explored in the context of catalysis and inverse electric catalysis~\cite{Loewe_Valenzuela_Zamora_2022}. In contrast, in this work we shall focus on the renormalization effects induced by the combination of electric field $\mathbf{E}$ and its stochastic fluctuations. For this purpose, we shall start by introducing the formalism to include the classical fluctuations of the background field into the Lagrangian, and the subsequent statistical averages by means of the replica trick. Later, we shall consider two different limits for the reference propagator, corresponding to: (a) The very weak electric field limit $|q\mathbf{E}|/m^2 \ll 1$ (for $q$ and $m$ the particle charge and mass), and the very strong field limit $|q\mathbf{E}|/m^2 \gg 1$, respectively. In both cases, we shall study the evolution of the self-interaction coupling as a function of the noise auto-correlation and the characteristic logarithmic momentum scale $t=\ln(p^2/m^2)$, by obtaining beta functions of the theory, and then solving the corresponding Callan-Symanzik equation for the renormalization group flows.\\

This article is organized as follows: In Sec.~\ref{sec2}, we introduce the modified $\lambda\phi^4$ Lagrangian with stochastic fluctuations. In Sec.~\ref{sec3}, we define the scalar boson propagator in the Schwinger proper-time representation for both limiting cases, which are then used in Sec.~\ref{sec4} to find the self-energy at one loop. Then, in Sec.~\ref{sec5} we define the physical parameters in the framework of renormalized perturbation theory, and find the corresponding self-energy counterterms. In Sec.~\ref{sec6}, we study the spectral density of the background and dressed propagators, in order to study the effects of the interaction on the physical mass. In Sec.~\ref{sec7}, we find the one-loop corrections to both interaction vertices, along with their corresponding counterterms. Finally, in Sec.~\ref{sec8} we study the renormalization group equations of the theory, in order to evaluate the evolution of both couplings with the energy scale. All detailed mathematical derivations are presented in Appendices~\ref{Ap_weak_field}--\ref{Ap_running_couplings}.


\section{The model}\label{sec2}\label{model}
We shall consider a system of self-interacting charged bosons whose dynamics is driven by the presence of a classical static electric field with stochastic spacetime fluctuations. Therefore, the corresponding background classical gauge field is composed by an average $A_{\rm{BG}}^{\mu}(x)$ and a stochastic fluctuation $\delta A_{\rm{BG}}^{\mu}(x)$ component, as follows
\begin{eqnarray}
A^{\mu}_{\rm{BG}}(x) + \delta A_{\rm{BG}}^{\mu}(x).
\end{eqnarray}
Here, we assume that the stochastic fluctuations satisfy the following "white-noise" statistical properties
\begin{align}
\left<\delta A^\mu_{\rm{BG}} (x)\right> &= 0,\\
\left<\delta A^\mu_{\rm{BG}} (x) \delta A_{\rm{BG}}^\nu (y) \right> &= \Delta \delta^{\mu\nu} \delta^4(x-y).\label{delta_def}
\end{align}
The subscript BG stands for ``background'', representing the external electromagnetic field imposed by experimental conditions. After Eq.~\eqref{delta_def}, the magnitude of the auto-correlation function for the fluctuations is defined as $\Delta$ which, as we will see later, will play the role of an effective coupling parameter between charged currents. The statistical properties of the stochastic fluctuations correspond to a Gaussian functional distribution of the form~\cite{Castano_Munoz1_2023}
\begin{align}
dP[\delta A_\text{BG}^\mu (x)] = \mathcal{N}\, e^{- \int d^4x \frac{\left(\delta A_\text{BG}^\mu(x)\right)^2}{2\Delta}} \mathcal{D}[\delta A_\text{BG}^\mu(x)] \label{distribution}.
\end{align}
For this purpose, we consider the standard $\lambda\phi^4$ theory for complex fields, where the fluctuating electric field is introduced, in the usual way, as a gauge connection in the covariant derivative
\begin{align}
\partial^\mu + iq\left(A^\mu_\text{BG}(x) + \delta A^\mu_\text{BG}(x)\right) \equiv D^{\mu} + iq \delta A^\mu_\text{BG}(x).
\end{align}
Here $q$ is the particle charge, $D^{\mu}$ is the covariant derivative with the connection $A^\mu_\text{BG}(x)$ corresponding to the average background gauge field, and $\delta A^\mu_\text{BG}(x)$ is the random fluctuation with respect to the mean value. 

The Lagrangian density for the model will be given by the sum of two contributions
\begin{align}
\mathcal{L} = \mathcal{L}_{BG} + \mathcal{L}_{\delta},
\end{align}
where the first one represents the standard $\lambda\phi^4$ theory coupled to a static electric field
\begin{align}
\mathcal{L}_{BG} =\left|D_{\mu}\phi(x)\right|^2 - m^2\phi^\dagger(x)& \phi(x)- \frac{\lambda}{4}\left[\phi^\dagger(x)\phi(x)\right]^2.
\end{align}
Here, $\phi$ is a complex scalar field, the operator $D_\mu$ represents the covariant derivative in the absence of fluctuations, and $\lambda>0$ is a coupling constant. The second contribution represents the interaction between the scalar field and the background fluctuations
\begin{align}
\mathcal{L}_\delta = - iq\delta A^\mu_\text{BG}(x)j_{\mu}(x).
\end{align}
\noindent where the currents will be given by
\begin{align}
j_{\mu}(x) = \phi^\dagger(x)D_\mu\phi(x) - \left(D_\mu\phi(x)\right)^\dagger\phi(x).\label{currents}
\end{align}

To determine the evolution of the physical parameters of the system, we must find the generating functional in the presence of the background electromagnetic field with stochastic fluctuations. The generating functional of the theory, in the absence of sources, and for a given realization of the background field and its fluctuations is given by the expression
\begin{align}
Z[A_\text{BG}] = \int \mathcal{D}\left[\phi^\dagger,\phi\right] e^{i\int d^4 x\, \left(\mathcal{L}_{BG} + \mathcal{L}_\delta\right)}.
\end{align}
In contrast with the quantum scalar field $\phi(x)$, the background fluctuations $\delta A^{\mu}_{BG}(x)$ are classical stochastic variables distributed according to Eq.~\eqref{distribution}. Since all the physical information is encoded in the averaged connected 2n-point quantum correlation functions, 
\bea
&&\overline{\langle \hat{T}\phi(x_1)\ldots\phi(x_n)\phi^{\dagger}(x_{n+1})\ldots\phi^{\dagger}(x_{2n}) \rangle}\nonumber\\
&&= \frac{(-i)^{2n}\delta^{2n}}{\delta J(x_
{2n})\ldots\delta J^{\dagger}(x_{1})} \overline{\ln \left(Z[A_{BG};J^{\dagger},J]\right)} |_{J=0}, 
\label{eq_avg_connected}
\eea
we must find the statistical average of $\ln Z$ over the background noise 
\begin{align}
\overline{\ln(Z[A_{BG}])} = \int dP[\delta A_{BG}] \ln\left(Z[A_{BG} + \delta A_{BG}]\right).
\end{align}
Computing $\overline{\ln Z}$ poses technical difficulties, which can be circumvented by the application of the replica trick~\cite{Mezard_1987}, following the mathematical identity 
\begin{align}
\overline{\ln\left(Z[A_{BG}]\right)} = \lim_{n\to 0} \frac{\overline{Z[A_{BG}]}-1}{n}.
\end{align}
Here $n$ represents the total number of replicas of the bosonic field. The method implies the introduction of $n$ replica fields $\phi_a(x)$, for $1\leq a \leq n$. This way, the average generating functional can be explicitly computed, leading to the expression
\begin{align}
\overline{Z^n} &= \int \prod_{a=1}^n \mathcal{D}\left[\phi^\dagger_a,\phi_a\right]dP[\delta A_\text{BG}^\mu(x)] e^{i\int d^4x \sum_{a=1}^n \left(\mathcal{L}_{BG} + \mathcal{L}_\delta\right)} \notag\\
&= \int \prod_{a=1}^n \mathcal{D}\left[\phi^\dagger_a,\phi_a\right] e^{i\int d^4 x\mathcal{L}_\text{eff}\left[\phi^\dagger_a,\phi_a\right]},\label{gen-fun}
\end{align}
where in the last step, we performed Gaussian functional integration over the probability distribution defined by Eq.~\eqref{distribution}. The resulting effective Lagrangian is thus given by the sum of two contributions
\begin{align}
\mathcal{L}_\text{eff} = \mathcal{L}_\text{BG} + \mathcal{L}_\text{int},
\end{align}
where 
\begin{align}
\mathcal{L}_\text{BG}&= \sum_{a} \left|D_{\mu}\phi_a(x)\right|^2 - m^2\phi^\dagger_a(x)\phi_a(x),\\[5pt]
\mathcal{L}_\text{int}&=- \frac{\lambda}{4}\sum_{a} \left[\phi^\dagger_a(x)\phi_a(x)\right]^2 - iq^2\Delta\sum_{a,b} j_{\mu,a}(x) j^{\mu,b}(x), \label{l-int}
\end{align}
represent the background and interaction Lagrangian, respectively. Notice that, for $d\rightarrow4$ the spacetime dimensions, the corresponding dimension of the currents defined by Eq.~\eqref{currents} is $m^{d-1}$, and hence the coupling $\Delta$ has dimensions of
\begin{align}
[\Delta]\ = m^{2-d}.
\end{align}
On the other hand, the parameter $\lambda$ has dimensions of \begin{align}
[\lambda]=m^{4-d}.
\end{align}
Hence, for consistency, introduce a factor of $m^2$ in the definition of the effective noise coupling, thus defining the dimensionless coupling (in $d=4$)
\begin{align}
\tilde{\Delta} = q^2m^2\Delta.
\end{align}

In summary, after averaging over the electromagnetic fluctuations, we are left with an effective theory describing a Klein-Gordon Lagrangian coupled to a constant background electromagnetic field, with a self-interaction $\lambda \phi^4$, and an effective interaction between charged currents, the latter proportional to the magnitude of the noise auto-correlation function $\Delta$. 


\section{Boson propagator in the presence of a uniform electric field}\label{sec3}\label{propagators}
To approach the phenomenological scenario of the in-plane electric field present in asymmetric heavy-ion collisions, we shall neglect the magnetic field, and consider a one-dimensional electric field in the $\hat{z}$-direction as follows
\begin{align}
\vec{E} &= E\,\hat{z},\\
\vec{B} &= \vec{0}.
\end{align}
These fields are incorporated into the definition of the electromagnetic tensor
$F^{\mu\nu} = \partial^{\mu}A^{\nu} - \partial^{\nu}A^{\mu}$. Using $-\partial_3 A^{0} = E$ (with other components equal to zero), the explicit form of the tensor is
\begin{align}
\hat{F} = \begin{pmatrix}
0 & 0 & 0 & i\, E\\
0 & 0 & 0 & 0\\
0 & 0 & 0 & 0\\
-i\,E & 0 & 0 & 0
\end{pmatrix},
\end{align}
which is written in Euclidean metric $A_0 \rightarrow i A_0$ for technical purposes. For a uniform background electric field, the Schwinger's proper-time representation for the free scalar propagator is~\cite{Schwinger_1951,dittrich1985effective}
\begin{align}
D_E(x,y) = \Phi(x,y) \int \frac{d^4p}{(2\pi)^4} e^{-ip(x-y)}D_E(p),
\end{align}
where $\Phi(x,y)$ is the Schwinger phase, defined on a straight line path $\xi^{\mu}(t)$, $t\in[0,1]$ connecting the points $x^{\mu} = \xi^{\mu}(1)$ and $y^{\mu}=\xi^{\mu}(0)$, respectively
\begin{align}
\Phi(x,y) = e^{-iq\int_y^x d\xi_\mu A^\mu_{\text{BG}}(\xi)}. \label{phase}
\end{align}
The translational invariant part of the propagator is expressed in momentum space $D_E(p)$, and for the case of a uniform electric field, it takes the form \cite{Loewe_Valenzuela_Zamora_2022,Loewe_Valenzuela_Zamora_2023}
\begin{align}
D_E(p) = \int_0^\infty ds\, \frac{e^{-s\left[\frac{\tanh(qEs)}{qEs}p_\parallel^2 + p_\perp^2 + m^2\right]}}{\cosh(qEs)},
\end{align}
where $s$ is the Schwinger proper time parameter, and and the subscript E stands for "electric". Here, the momentum 
\begin{align}
p^2 &= p_\parallel^2 + p_\perp^2
\end{align}
is in Euclidean metric, with parallel and transverse components defined by
\begin{align}
p_\parallel^2 &= p_4^2 + p_3^2,\\
p_\perp^2 &= p_1^2 + p_2^2.
\end{align}

\subsection{Propagator in the weak electric field regime}

Restricting ourselves to the weak electric field regime ($|qE| \ll m^2$), the background propagator is given by the series expansion (see Appendix~\ref{Ap_weak_field} for details)
\begin{align}
D_E(p) = \frac{1}{p^2+m^2} + \left(qE\right)^2\left[\frac{-1}{\left(p^2+m^2\right)^3}+\frac{2p_\parallel^2}{\left(p^2+m^2\right)^4}\right]. \label{weak-prop}
\end{align}

As both contributions to the interacting Lagrangian from Eq.~\eqref{l-int} are added separately, regular procedures of perturbation theory can be carried out for both interactions without further complications. In particular, we shall obtain the self-energy and the vertex corrections at first and second order in the couplings $\lambda$ and $\tilde{\Delta}$, respectively.

\subsection{Propagator in the strong electric field regime}
A more general representation of the scalar propagator for a finite electric field is given by the infinite series~\cite{Loewe_Valenzuela_Zamora_2022}
\begin{eqnarray}
D_E(p) = 2\sum_{l=0}^{\infty}\left(-1\right)^l \frac{e^{-\frac{p_{\parallel}^2}{q E}}L_{l}(2 p_{\parallel}^2/(qE))}{p_{\perp}^2 + (2 l +1)q E + m^2}.
\label{eq_pro_full}
\end{eqnarray}
The gap between the lowest level $l=0$, and the first excited level $l = 1$ grows with the magnitude of the electric field. Therefore, in the very strong field limit $|q E|/m^2 \gg 1$, the propagator will be dominated by the $l=0$ mode, and hence
\begin{eqnarray}
D_E(p) \approx 2 \frac{e^{-\frac{p_{\parallel}^2}{q E}}}{p_{\perp}^2 +m_E^2},
\label{strong-prop}
\end{eqnarray}
where
\begin{align}
m_E^2 = qE + m^2
\label{eq_mesquare}
\end{align}
is defined as an effective electric mass. 

In what follows, we shall consider the renormalization of the theory, in the presence of the electromagnetic noise, for both regimes in the magnitude of the average background electric field, i.e.: (a) Very weak field $|q E|/m^2 \ll 1$, and (b) Very strong field $|q E| /m^2 \gg 1$, respectively.


\section{Self-energy at first order}\label{sec4}\label{s-energy}
Expanding the averaged generating functional from Eq.~\eqref{gen-fun} up to first order in the interaction Lagrangian, the two-point, time-ordered correlation function averaged over fluctuations, as defined in Eq.~\eqref{eq_avg_connected}, is given by the expression
\begin{align}
\overline{\left<\hat{T}\phi_\alpha(x)\phi_\beta^\dagger(y)\right>} &= \int\prod_{a=1}^n \mathcal{D}[\phi^\dagger_a(z),\phi_a(z)]e^{i\int d^4z \mathcal{L}_\text{BG}} \notag\\
&\hspace{25pt}\times\left(1+i\int d^4z \mathcal{L}_\text{int}\right)\phi_\alpha(x)\phi^\dagger_\beta(y).
\label{eq_twopoint}
\end{align}
Here, the propagator for the reference system, i.e. a non-interacting scalar theory in a constant background electric field, is given by the Schwinger propagator discussed in the previous section
\begin{align}
\left<\hat{T}\phi_{\alpha}(x)\phi_{\beta}^\dagger (y)\right>_\text{BG} &= \delta_{\alpha\beta}\, i D_E(x-y)\notag\\
&= i\delta_{\alpha\beta}\Phi(x,y) \int \frac{d^4p}{(2\pi)^4} e^{ip(x-y)} D_E(p),
\label{eq_propSchwinger}
\end{align}
which is $\propto\delta_{\alpha\beta}$, and hence explicitly diagonal in the replica component space.

As $\mathcal{L}_\text{int}$ contains the contributions from both types of interaction, proportional to the couplings $\lambda$ and $\tilde{\Delta}$, respectively, the averaged dressed propagator 
\begin{equation}
\overline{\langle \hat{T}\phi_{\alpha}(x)\phi_{\beta}^{\dagger}(y) \rangle} = i\delta_{\alpha\beta}\Phi(x,y)\int\frac{d^4 p}{(2\pi)^4} e^{i p\cdot(x-y)}D_{\text{int}}(p)
\end{equation}
satisfies a Dyson equation. Using the property of the Schwinger phase $\Phi(x,z)\Phi(z,y) = \Phi(x,y)$, the Dyson equation in momentum space is given by
\begin{align}
iD_\text{int}(p) &= iD_E(p)+iD_E(p)\left[-i\Sigma(p)\right]iD_\text{int}(p).\label{dyson}
\end{align}
Here, the self-energy is given by the sum of two contributions: the first one corresponds to the $\lambda$-interaction,  while the second one is associated to the effective interaction ($\tilde{\Delta}$-interaction) induced after averaging over background electric fluctuations, 
\begin{align}
\Sigma(p,E) = \Sigma_\lambda(p,E) + \Sigma_\Delta(p,E).
\end{align}
Notice that the self energy has an explicit dependence on the background electric field $E=|\mathbf{E}|$, which will become relevant when we renormalize the theory.\\

\subsection{First-order in $\lambda$ contribution to the self energy}

We now show that the $\lambda$-contribution to the self energy is obtained in a straightforward way. After Eq.~\eqref{eq_twopoint}, the first order in $\lambda$ correction to the averaged propagator is given by the expression
\begin{align}
&\hspace{-5pt}\overline{\left<\hat{T}\phi_\alpha(x)\phi^\dagger_\beta(y)\right>}_\lambda = \notag\\
&\hspace{5pt}-\frac{i\lambda}{4} \int d^4z \left<\hat{T}\phi^\dagger_a(z)\phi_a(z)\phi^\dagger_b(z)\phi_b(z)\phi_\alpha(x)\phi^\dagger_\beta(y)\right>_\text{BG}, \label{self-lambda}
\end{align}
where the expectation value of the field product is defined for a reference system given by the Lagrangian with a constant background electric field $\mathcal{L}_\text{BG}$, as defined in Eq.~\eqref{l-int}. 
Then, applying Wick's theorem, the contractions of the fields are factored as usual, where each contraction is expressed in terms of the reference system Schwinger propagator Eq.~\eqref{eq_propSchwinger}.

\begin{figure}[H]
\centering
\includegraphics[width=0.3\linewidth]{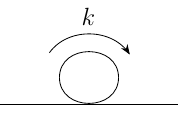}
\caption{Feynman diagram for the first order in $\lambda$-correction to the propagator.}
\label{1ordercorr}
\end{figure}

Following this procedure for each contraction, and integrating over $z$, we find that the self-energy contribution arising from the $\lambda$-interaction is given by
\begin{align}
\Sigma_\lambda(p,E) = -\frac{\lambda}{2} \int \frac{d^4k}{(2\pi)^4}  D_E(k),
\label{self-l}
\end{align}
which is independent of the external momentum and corresponds to the typical one loop correction for $\lambda\phi^4$ theory shown in Figure~\ref{1ordercorr}. The difference in the present case lies in the fact that the propagator is now dressed by the effect of the background electric field, as defined in Eq.~\eqref{weak-prop}.

\subsection{First-order in $\Delta$ contribution to the self-energy}

\begin{figure}[t]
\centering
\includegraphics[width=0.55\linewidth]{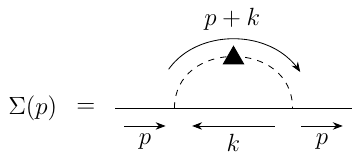}
\caption{Self energy diagram at first order in $\tilde{\Delta}$.}
\label{self-energy}
\end{figure}

For the effective interaction induced by the statistical average over electromagnetic fluctuations, the procedure is very similar to the $\lambda$ case already discussed, but we must also consider the effect of the covariant derivative $D_\mu (x) = \partial_\mu + iq A^\mu_\text{BG}(x)$ on the currents defined by Eq.~\eqref{currents}. The action of this operator over the propagator is as follows
\begin{align}
D_\mu(x)\, iD_E(x-y) = \Phi(x,y) \int \frac{d^4p}{(2\pi)^4}e^{ip(x-y)}\,p\cdot iD_E(p).
\end{align}
Taking this property into account, and following a similar procedure as in the previous case, the self-energy contribution due to averaging over background electromagnetic fluctuations is depicted as a Feynman diagram in Fig.~\ref{self-energy}, and given by the analytical expression
\begin{align}
\Sigma_\Delta(p,E) = -\frac{2i\tilde{\Delta}}{m^2}  \int \frac{d^4k}{(2\pi)^4}\,D_E(k)\left(p-k\right)^2.\label{self-delta}
\end{align}
By joining the two contributions at first order in $\lambda$ and $\tilde{\Delta}$, respectively, we notice that the total self energy is a complex quantity, with its real and imaginary parts given by
\begin{align}
\Re[\Sigma] &= \Sigma_\lambda(p,E),\\
\Im[\Sigma] &= \Sigma_\Delta(p,E).
\label{self_real_Im}
\end{align}

\section{Self energy counterterms and renormalized parameters}\label{sec5}\label{counterterms}
The self energy contributions defined by Eq.~\eqref{self-l} and Eq.~\eqref{self-delta}, respectively, are computed in detail in Appendix~\ref{AppSelfEnergy}. As a result, we find that the electric field-independent contributions to the self-energy will result in logarithmically divergent integrals, for which we use dimensional regularization (see Appendix~\ref{AppSelfEnergy} for details). Then, we must proceed in the framework of renormalized perturbation theory in order to renormalize the mass and couplings in the Lagrangian.\\

We first start by rescaling the bare field, introducing the renormalization factor $Z$ as
\begin{align}
\phi_a^{(0)}(x) = \sqrt{Z} \phi_a(x),
\end{align}
where $\phi_a(x)$ stands for the renormalized field components. Then, the rescaled field is introduced in the Lagrangian, which is rewritten in terms of the bare ($m_0$, $\lambda_0$, $\tilde{\Delta}_0$) and physical parameters ($m^2,\,\tilde{\Delta},\,\lambda$), as shown in Appendix \ref{renorm-lag}. In particular, we define the following counterterms for the mass, field strength, and couplings, respectively
\begin{align}
&\delta_Z = Z-1\,,\label{delta-z}\\
&\delta_{m^2}= m_0^2 Z - m^2\,,\\
&\delta_\lambda = \lambda_0 Z^2 - \lambda,\label{delta-lambda}\\
&\delta_{\tilde{\Delta}} = {\tilde{\Delta}}_0 Z^2 - {\tilde{\Delta}} \label{delta-delta}.
\end{align}
Taking these counterterms into account, the renormalized self-energy is given, up to one-loop, by the expression in Euclidean metric
\begin{align}
-i\Sigma^R(p,E) = -i\Sigma(p,E) - i\left(p^2\delta_Z + \delta_{m^2}\right).
\label{eq_selfR}
\end{align}
We define the renormalization conditions as follows
\begin{eqnarray}
\lim_{p^2 \rightarrow m^2} \Sigma^R(p,E) &=& \lim_{p^2 \rightarrow m^2}\Sigma(p,E) - \left( m^2\delta_Z + \delta_{m^2} \right)\nonumber\\ 
&=& 0,\label{sigma-cond}
\end{eqnarray}
and
\begin{eqnarray}
\lim_{p^2 \rightarrow m^2} \frac{\partial}{\partial p^2}\Sigma^R(p,E) &=& \lim_{p^2 \rightarrow m^2}\frac{\partial}{\partial p^2}\Sigma(p,E) - \delta_Z\nonumber\\[5pt]
&=&  0. \label{z-cond}
\end{eqnarray}

\subsection{Weak field limit $|qE|/m^2\ll 1$}
As shown in Appendix~\ref{AppSelfEnergy}, the total self-energy at first-order in the couplings $\lambda$ and $\tilde{\Delta}$ is given by 
\begin{align}
\Sigma(p,E) =& \frac{1}{(4\pi)^2}\left[\frac{\lambda}{2}+\frac{2i\tilde{\Delta}}{m^2}\left(p^2-m^2\right)\right] \notag\\
&\times \left\{m^2\left[\frac{2}{\epsilon} - \gamma + \ln\left(\frac{4\pi\mu^2}{m^2}\right) + 1 \right] + \frac{(qE)^2}{6m^2} \right\},
\end{align}
where $\epsilon = 4 - d$ is the dimensional regularization parameter, and $\mu$ is the renormalization energy scale.\\

Solving the system formed by Eqs.~\eqref{sigma-cond} and \eqref{z-cond} explicitly, we find the counterterms for the mass and the field strength, respectively, in the weak field regime as follows
\begin{align}
\delta_Z &= -\frac{2i\tilde{\Delta}}{(4\pi)^2} \left[\frac{2}{\epsilon}-\gamma + \ln\left(\frac{4\pi\mu^2}{m^2}\right) +1 \right]\label{dzw},\\[5pt]
\delta_{m^2} &= \frac{m^2}{(4\pi)^2}\left[\frac{2}{\epsilon}-\gamma + \ln\left(\frac{4\pi\mu^2}{m^2}\right) +1 \right]\left[\frac{\lambda}{2}+2i\tilde{\Delta}\right].
\end{align}
We notice that both counterterms are complex due to the electromagnetic noise contribution $i\tilde{\Delta}$, and according to the definition of the renormalized parameters from Eqs.~\eqref{delta-z}-\eqref{delta-delta} this will give rise to complex renormalized couplings.

\subsection{Very strong field regime $|qE|/m^2\gg 1$}
As shown in detail in the Appendix~\ref{AppSelfEnergy}, the self-energy in the very strong field limit is given by the expression
\begin{eqnarray}
\Sigma(p,E) &=& \frac{1}{(4\pi)^2}\frac{(qE)}{m^2}\left[4i \tilde{\Delta} (p^2 - m^2) - m^2\lambda 
\right]\nonumber\\
&&\hspace{1cm}\times\left[\frac{2}{\epsilon} - \gamma + \ln\left( \frac{4\pi\mu^2}{m^2_E} \right)  \right],\label{self-strong}
\end{eqnarray}
with $m_E^2$ as defined in Eq.~\eqref{eq_mesquare}.
In this case, we also define the renormalized self-energy according to Eq.~\eqref{eq_selfR}. Then, after applying the renormalization conditions defined in Eq.~\eqref{sigma-cond} and Eq.~\eqref{z-cond}, we obtain 
explicit expressions for the counterterms in the very strong field limit $|qE|/m^2\gg 1$
\begin{align}
\delta_Z &= -\frac{4i\tilde{\Delta}}{(4\pi)^2}\frac{(qE)}{m^2} \left[\frac{2}{\epsilon} - \gamma + \ln\left( \frac{4\pi\mu^2}{m^2_E} \right)  \right],\\[5pt]
\delta_{m^2} &= \frac{(qE)}{(4\pi)^2}\left[\lambda+4i\tilde{\Delta}\right] \left[\frac{2}{\epsilon} - \gamma + \ln\left( \frac{4\pi\mu^2}{m^2_E} \right)\right].\label{dzs}
\end{align}



\section{Spectral density analysis}\label{sec-6}\label{sec6}
To analyze the effects of the background and noisy electric fields over the physical mass, we must compute the spectral densities of the theory. The dressed propagator is defined by the Dyson equation~\eqref{dyson}, where the self-energy will have real and imaginary contributions as defined in Eq.~\eqref{self_real_Im}. The spectral density will be defined by the means of the Källén-Lehmann representation of the two point function as
\begin{align}
\rho_{\text{int}}\left(p\right) = -\frac{1}{\pi}\Im[D_{\text{int}}(p)]. \label{kl-rep}
\end{align}
This definition is valid in Minkowski space, and hence we revert the propagators and self-energy contributions to Minkowski metric by performing a Wick rotation on the 4-momentum as $p_4 \to -ip_0$. Notice that this changes the definition of the parallel component of the momentum as
\begin{align}
p_\parallel^2 \to -p_0^2 + p_3^2.
\end{align}
We will also compare the effects of the interaction with the reference system, for which we will define the spectral density 
\begin{align}
\rho_E\left(p\right) = -\frac{1}{\pi}\Im[D_E(p)].
\end{align}
Here, $D_E(p)$ corresponds to the reference system propagator, either for the weak, Eq.~\eqref{weak-prop}, or the strong, Eq.~\eqref{strong-prop}, electric field cases, respectively.

\subsection{Weak electric field regime $|qE|/m^2 \ll 1$}
In this case, the renormalized self-energy is given by
\begin{align}
\Sigma^{R}(p,E) =& \frac{1}{(4\pi)^2}\frac{(qE)^2}{6m^2}\left[\frac{\lambda}{2}+\frac{2i\tilde{\Delta}}{m^2}\left(p^2-m^2\right)\right].
\end{align}

\begin{figure}[t]
\centering
\includegraphics[width=\linewidth]{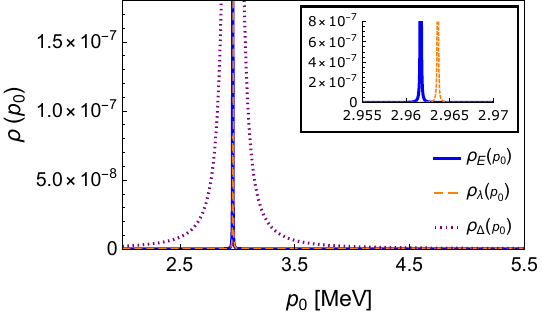}
\caption{Spectral density for the bare and dressed propagators for fixed values of: mass ($m_\pi=140$ MeV), couplings ($\lambda=2.06$, $\tilde{\Delta}=5\times10^{-7}$) and electric field $\left(|qE|=10^{-6}\right.$ MeV$\left.^2\right)$.}
\label{rho-weak}
\end{figure}

The corresponding spectral density for the dressed propagator is shown in Figure~\ref{rho-weak}, as a function of the momentum scale $p_0$, considering for the sake of the example the numerical parameters $\lambda = 2,06$, $\tilde{\Delta} = 5\times 10^{-7}$, the pion mass $m_{\pi} = 140\,\text{MeV}$, and an average background electric field $|q E| = 10^{-6}\,\text{MeV}^2$. The spectral density plot shows the emergence of quasi-particle states, with well-defined mass and lifetime, the later being inversely proportional to $\tilde{\Delta}$.\\

The real contribution to the self-energy slightly shifts the resonance to the right, as shown by the inset plot in Figure~\ref{rho-weak}, meaning that the pole mass is increased by the effect of the self-interaction. On the other hand, the effect of the imaginary part of the self-energy, given by the fluctuations, is observed by the spectral broadening of the resonance, thus taking the form of a Lorentzian distribution. This resonance implies the emergence of quasi-particle states, where the width of the Lorentzian is inversely proportional to the lifetime of these states, i.e. inversely proportional to the noise auto-correlation $\tilde{\Delta}$.\\


\subsection{Very strong electric field regime $|qE|/m^2 \gg 1$}
In the very strong field regime $|qE|/m^2 \gg 1$, the self-energy given by Eq.~\eqref{self-strong} is UV-logarithmically divergent with no finite contributions. Therefore, after adding the counterterms $\delta_Z$ and $\delta_{m^2}$ to the renormalized Lagrangian, the corresponding renormalized self-energy vanishes. Therefore, the spectral density of the dressed propagator will not receive contributions from neither interaction at this order.\\

The only finite spectral density function will be given by $\rho_E(p)$. Notice that the form of the propagator in Eq.~\eqref{strong-prop} cannot be described using the Källén-Lehmann representation. Nonetheless, if we evaluate the spectral density at $p_\parallel^2=0$, then the propagator takes a more conventional form, which in Euclidean metric results
\begin{align}
D_E\left(p_\perp\right)\xrightarrow{p_\parallel^2=0} \frac{1}{p_\perp^2 + m_E^2},
\end{align}
and then, the spectral density can be represented using Eq. \eqref{kl-rep}. By the form of the propagator, we can see that the pole mass will be only modified by the value of the electric field, as defined in Eq.~\eqref{eq_mesquare}. As the magnitude of the field increases, the spectral density will approach that of a free-particle, which is consistent with a theory that exhibits asymptotic freedom, an issue that will be discussed further in Sec.~\ref{strong-rge}.

\section{Vertex corrections up to second order}\label{sec7}
As we expand the four point function up to second order in the interaction Lagrangian, we find 3 contributions: one for each separate coupling $\lambda$ and $\tilde{\Delta}$, respectively, and a mixing term proportional to both of them. As we are interested in a comparison of both interaction effects, for the sake of simplicity we shall ignore the mixing term in our present analysis.\\

\begin{figure}[H]
\centering
\includegraphics[width=\linewidth]{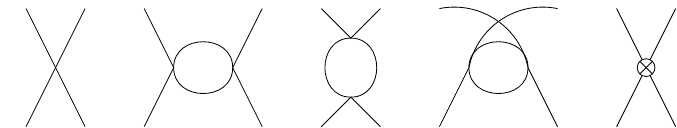}
\caption{Diagrams contributing to the 4-point vertex amplitude.}
\label{4p-amp}
\end{figure}

The vertex amplitude for both couplings is given by the similar diagrams as for the usual $\lambda\phi^4$ theory, as shown in Fig.~\ref{4p-amp}. In particular, the correction to the vertex is given by the sum of all $s, t$ and $u$ channels, such that we define the vertex amplitude for the coupling $g = (\lambda,\tilde{\Delta})$ as
\begin{align}
i\mathcal{M}_g(p_1p_2\to p_3p_4) &=\notag\\
&\hspace{-70pt}-ig + (-ig)^2\left[iV_g(4m^2,E)+2iV_g(0,E)\right]+ i\delta_g.\label{vertex-amp}
\end{align}
Here, $V_g(p,E)$ represents the vertex function, to be defined later for each coupling explicitly, and $\delta_g$ is the corresponding counterterm at the given order.

\begin{figure}[H]
\centering
\includegraphics[width=0.65\linewidth]{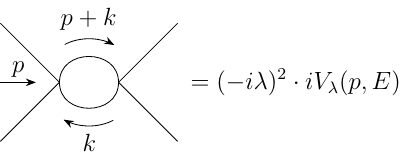}
\caption{$s-$channel contribution to the self interaction vertex correction at second order in $\lambda$.}
\label{lambda-s-channel}
\end{figure}

The $s-$channel diagram contributing to the vertex correction at second order in $\lambda$ is shown in Fig. \ref{lambda-s-channel}. 

\begin{figure}[H]
\centering
\includegraphics[width=0.7\linewidth]{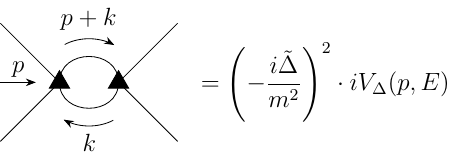}
\caption{s-channel diagram for the  $\tilde{\Delta}$-vertex at second order.}
\label{fig_s_Delta}
\end{figure}

Regarding the $\tilde{\Delta}$ vertex, we must adapt the general structure of the Feynman diagrams in Fig.~\ref{self-energy} to obtain the corresponding vertex rule.
\begin{figure}[H]
\centering
\includegraphics[width=0.65\linewidth]{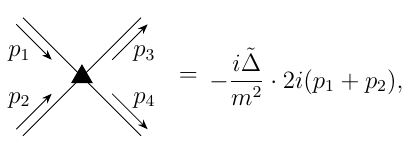}
\caption{Feynman rule for the $\tilde{\Delta}$-vertex at tree-level.}
\label{fig_vertex_Delta}
\end{figure}
\noindent Here, momentum conservation condition $p_1+p_2=p_3+p_4$ is satisfied. Then, at each vertex we have a factor of $(2ip)$, where $p$ is the incoming momentum. Using the Feynman rule for the interaction, the s-channel of the vertex correction is depicted in the Feynman diagram in Fig.~\ref{fig_s_Delta}.

\begin{figure}[H]
    \centering
    \includegraphics[width=0.65\linewidth]{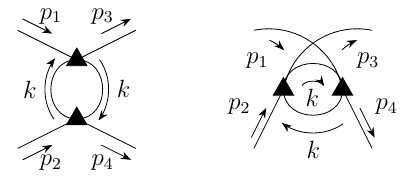}
    \caption{$t$ and $u-$channel diagrams for the one-loop correction to the $\tilde{\Delta}$-vertex.}
    \label{t-u-channels}
\end{figure}

while the $t$ and $u$ channels are depicted in the Feynman diagrams in Fig.~\ref{t-u-channels}.

Finally, the vertex functions for both contributions are given by 
\begin{align}
iV_\lambda(p,E) &= -\frac{i}{2}\cdot  iV\left(p,E\right),\\[5pt]
iV_{\Delta}(p,E) &= -4ip^2 \cdot iV\left(p,E\right),
\end{align}
where all expressions are in Euclidean metric. In particular, for the s-channel diagram, we define the generalized vertex function as
\begin{align}
iV(p,E) = {\int \frac{d^4k}{(2\pi)^4}}\,iD_E(k)\,iD_E(p+k),
\end{align}
whose explicit form will depend on the specialized approximation for the propagator used for each electric field regime.

\subsection{Weak field regime $|qE|/m^2 \ll 1$}
As we are interested in obtaining the beta functions for the renormalized parameters in the theory, we only need to compute the divergent terms of the vertex function, which are independent of the background electric field. In particular, this contribution is given by the integral
\begin{align}
iV(p,0) = \int\frac{d^4k}{(2\pi)^4} \frac{1}{\left[(p+k)^2+m^2\right](k^2+m^2)}.
\end{align}
As shown in detail in Appendix~\ref{AppVertex}, this integral is computed using the standard technique of Feynman parameters and dimensional regularization, resulting in
\begin{align}
iV(p,0) = \frac{1}{(4\pi)^2}\left[\frac{2}{\epsilon}-\gamma+\ln\left(\frac{4\pi\mu^2}{m^2}\right)\right]+\text{finite terms},
\end{align}
where $\epsilon = 4 - d$ and the finite contribution corresponds to terms regular in $p$. 

\subsection{Very strong field regime $|qE|/m^2 \gg 1$}
The procedure to compute the vertex function is shown in Appendix~\ref{AppVertex}, resulting in
\begin{align}
iV(p,E) = \frac{qE}{2(4\pi p_\perp)^2}\frac{e^{-\frac{p_\parallel^2}{2qE}}}{(x_2-x_1)}\ln\left(\frac{1-x_2}{1-x_1}\cdot\frac{x_1}{x_2}\right),
\end{align}
where
\begin{align}
x_{1,2}&= \frac{1}{2} \pm \sqrt{1+\frac{4m_E^2}{p_\perp^2}}.
\end{align}
We remark that the vertex correction in the strong field case is free of divergences, and hence it does not contribute to the renormalization of the couplings, at least up to one loop.

\subsection{Vertex counterterms}
From the previous results, we notice that the vertex amplitude is UV-logarithmically divergent in the weak electric field regime, but it is free of divergences in the very strong field regime. As in the case of the self energy, the divergent integrals for the vertex function are independent of the electric field. Recalling the definition of the vertex amplitude from Eq.~\eqref{vertex-amp}, the vertex counterterms for the couplings $g = (\lambda,\tilde{\Delta})$ are defined by
\begin{align}
i\delta_g = (-ig)^2\left[iV_g(4m^2,E=0)+2iV_g(0,E=0)\right].
\end{align}

Using this definition and the divergent contribution to the vertex functions, we find the vertex counterterms in the weak electric field regime $|qE|/m^2\ll 1$ for both couplings as follows
\begin{align}
\delta_{\lambda} &= -\frac{3\lambda^2}{32\pi^2}\left[\frac{2}{\epsilon}-\gamma+\ln\left(\frac{4\pi\mu^2}{m^2}\right)\right], \label{dlw}\\[5pt]
\delta_{\tilde{\Delta}} &= -\frac{\tilde{\Delta}^2}{\pi^2}\left[\frac{2}{\epsilon}-\gamma+\ln\left(\frac{4\pi\mu^2}{m^2}\right)\right].\label{ddw}
\end{align}

In contrast, in the very strong electric field regime $|q E|/m^2 \gg 1$ we have
\begin{align}
\delta_\lambda = \delta_{\tilde{\Delta}} = 0.
\end{align}



\section{Renormalization group equations}\label{sec8}
\begin{figure*}[t]
\begin{minipage}{0.47\linewidth}
\centering
\includegraphics[width=0.9\linewidth]{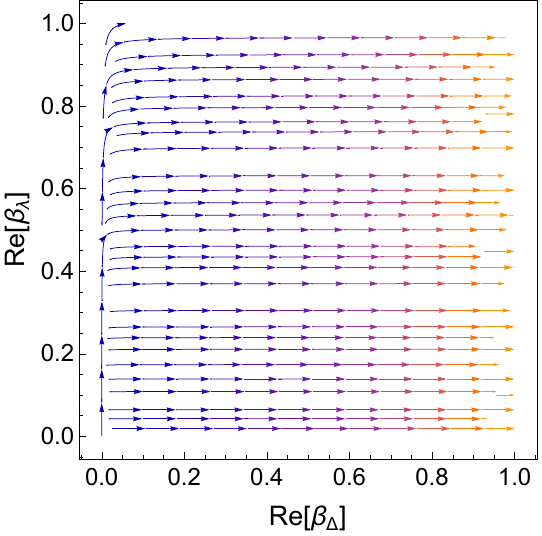}
\captionof{subfigure}{Real part streamplot.}
\label{re-stream}
\end{minipage}
\hspace{5pt}
\begin{minipage}{0.47\linewidth}
\centering
\includegraphics[width=0.9\linewidth]{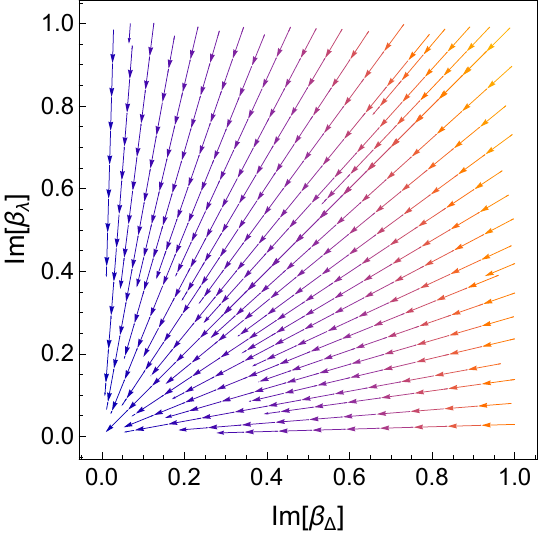}
\captionof{subfigure}{Imaginary part streamplot.}
\label{im-stream}
\end{minipage}
\caption{Stream plot of the beta functions in the very weak electric field regime $|qE|/m^2\ll 1$.}
\label{stream}
\end{figure*}
As sown in the previous section, the vertex functions renormalize very differently depending on the electric field intensity. In the weak electric field regime, the vertex function exhibits UV-logarithmic divergences, which are absorbed in suitable counterterms, while in the very strong electric field regime the function is free of divergences, at least up to one loop order. Nonetheless, the couplings will still flow with the logarithmic momentum scale, as both couplings also depend on the field-strength renormalization factor $Z$, after the definition in Eqs.~\eqref{delta-lambda} and~\eqref{delta-delta}. Therefore, for both regimes we formulate the renormalization group equations (RGE), in order to find the running couplings and their dependence on the logarithmic momentum scale. We define the beta function for each parameter $g = (\tilde{\Delta},\lambda)$ as
\begin{align}
\beta(g) &= \frac{d g}{d\ln\mu}.
\end{align}
Then, considering the definition of the counterterms, the RGE for the couplings is
\begin{align}
\beta_g&=\frac{d}{d\ln\mu}\left(g_0Z^2-\delta_g\right)\notag\\[5pt]
&=\frac{d}{d\ln\mu}\left[ 2 g_0 \delta_Z -\delta_g\right].
\end{align}
In this section, we will study the RGE flow of the system of $\beta$ functions, along with the solution of the system in both the weak and strong electric field regimes, respectively. 

\subsection{Weak electric field regime}
Using the counterterms for the vertex from Eqs.~\eqref{dlw} and \eqref{ddw}, and the counterterm for the field strength factor from Eq. \eqref{dzw} we find the beta functions for the renormalized couplings $\lambda=\lambda(t)$ and $\tilde{\Delta}=\tilde{\Delta}(t)$ to be given by
\begin{align}
&\beta_{\tilde{\Delta}} = \frac{2}{\pi^2}\tilde{\Delta}^2\left(1-\frac{i}{4}\right)\label{bd},\\[5pt]
&\beta_\lambda = \frac{3}{16\pi^2}\lambda^2 - \frac{i}{2\pi^2}\lambda\tilde{\Delta},\label{bl}
\end{align}
where we defined the logarithmic momentum scale
\begin{align}
t = \frac{1}{2}\ln\left(\frac{p^2}{\mu^2}\right).
\end{align}
Before finding the solution, we can study the competition between both couplings by the means of the stream plots shown in Figs.~\ref{stream}-(a,b). This panel of figures represents a vector field with components $(\beta_{\tilde{\Delta}},\beta_\lambda)$, where the arrows indicate the direction of the RGE flow as a function of $t = \frac{1}{2}\ln(p^2/\mu^2)$, for the real (subfigure~\ref{stream}-(a)) and imaginary (subfigure~\ref{stream}-(b)) parts, respectively. Notice that the scale of $\tilde{\Delta}$ has been amplified for illustration purposes. As seen in Fig.~\ref{stream}-(a), the real parts of both couplings grow with the logarithmic momentum scale $t$, with $\lambda$ growing faster than $\tilde{\Delta}$. The noise contribution works as a damping factor for the self interaction, with $\tilde{\Delta}$ remaining almost constant as the energy scale increases. 

The imaginary part of the stream plot, displayed in Fig.~\ref{stream}-(b) shows the existence of a UV-stable fixed point at ${\text{Im}}\tilde{\Delta}={\text{Im}}\lambda=0$, thus implying that the imaginary part of both couplings becomes negligible at large energy scales.\\

Regarding the solution of the renormalization group equations, Eqs. \eqref{bd} and \eqref{bl} constitute a set of three coupled non-linear differential equations that can be solved analytically, as shown in detail in Appendix~\ref{Ap_running_couplings}. 
For the running effective coupling related to the electric noise, we obtain
\begin{eqnarray}
\tilde{\Delta}(t) = \frac{\tilde{\Delta}_0}{1 - \frac{2}{\pi^2}\left( 1 - \frac{i}{4} \right)\tilde{\Delta}_0 t}. 
\label{eq_Deltat}
\end{eqnarray}
The real part of the denominator defines a first Landau pole in the logarithmic momentum scale, such that the solution is applicable for $t < \pi^2/(2\tilde{\Delta}_0)$.

On the other hand, the self-interaction running coupling $\lambda(t)$ evolves as a function of the logarithmic momentum scale as follows
\bea
\lambda(t) 
&=& \frac{\lambda_0 \left( \tilde{\Delta}(t)/\tilde{\Delta}_0  \right)^{\frac{i}{i-4}}}{1 - \frac{3\lambda_0}{16\pi^2}\frac{\pi^2}{2\tilde{\Delta}_0}\left[ 1 - \left( \tilde{\Delta}(t)/\tilde{\Delta}_0  \right)^{\frac{4}{i-4}}  \right]}.
\label{eq_lambdat}
\eea
It is important to remark that, in the noiseless limit $\tilde{\Delta}_0\rightarrow 0$, the expression for the running self-interacting coupling reduces to the well known result
\begin{eqnarray}
\lim_{\tilde{\Delta}_0\rightarrow 0}\lambda(t) = \frac{\lambda_0}{1 - \frac{3\lambda_0}{16\pi^2}t},
\end{eqnarray}
which is valid for the logarithmic momentum scale well below the Landau pole, i.e. $t < 16\pi^2/(3 \lambda_0)$. Taking into account both Landau poles, we have that the analysis is valid for 
\begin{equation}
t < \min\left\{\frac{\pi^2}{(2\tilde{\Delta}_0)},\frac{3\lambda_0}{(16\pi^2)} \right\}.
\end{equation}

\subsection{Very strong electric field regime}\label{strong-rge}
Considerig the counterterm for the field strength factor in Eq. \eqref{dzs}, the beta functions for the renormalized couplings are given by
\begin{align}
\beta_{\tilde{\Delta}} &= -i\frac{qE}{\pi^2m^2}\tilde{\Delta}^2\label{bd-strong},\\[5pt]
\beta_\lambda &= -i\frac{qE}{\pi^2m^2} \lambda \tilde{\Delta}\label{bl-strong}.
\end{align}
The dynamics of both equations are somewhat trivial, and the RGE flow behaves the same as the imaginary part of the RGE flow in the very weak electric field case, represented by Fig.~\ref{stream}-(b).\\

Eqs.~\eqref{bd-strong} and \eqref{bl-strong} can be easily solved by separation of variables, as shown in Appendix~\ref{strong-running}. Then, the evolution of both couplings as a function of the logarithmic momentum scale is given by the expressions
\begin{align}
\tilde{\Delta}(t) &= \frac{\tilde{\Delta}_0}{1 + i\frac{qE}{\pi^2m^2}\tilde{\Delta}_0 t},\\[5pt]
\lambda(t) &= \frac{\lambda_0}{\tilde{\Delta}_0}\tilde{\Delta}(t).
\end{align}
We remark that in the noiseless limit, the bare self interaction coupling is recovered, as expected
\begin{align}
\lim_{\tilde{\Delta}_0\to 0} \lambda(t) = \lambda_0.  
\end{align}
Furthermore, both expressions lack the presence of a Landau pole, as opposed to the weak electric field case. Then, both couplings grow as $\sim 1/t$ with the energy scale, which is consistent with the emergence of asymptotic freedom at very large values of the electric field.

\section{Conclusions}
In this work, we have studied the effects of a classical electric field background with stochastic fluctuations over a self-interacting theory of complex scalar fields. Although a toy model, it represents a technically tractable effective field theory for pions under non-trivial background conditions, such as very weak ($|qE|/m^2\ll1$), or very strong ($|qE|/m^2\gg1$) average electric fields, respectively, subjected to stochastic fluctuations. This configuration is relevant in a variety of physical scenarios, particularly in heavy-ion collisions, where an in-plane electric field emerges due to nuclear charge anisotropies, specially when asymmetric nuclei are involved. 

The background electric field was introduced in the Lagrangian density via minimal coupling, as a connection in the covariant derivative. The space-time anisotropies in the background field are modeled as white noise, meaning the stochastic fluctuations satisfy a Gaussian functional distribution. Starting from the usual, complex $\lambda\phi^4$ Lagrangian, where the covariant derivative was modified to introduce the electric field with stochastic fluctuations $A^\mu_\text{BG} + \delta A^\mu_\text{BG}$, we followed the procedure established by some of us in previous works~\cite{Castano_Munoz1_2023,Castano_Munoz1_2024,Castano_Munoz2_2024,Castano_Munoz_2025}: to calculate the statistical average of the generating functional, given the many different realizations of the Lagrangian for each possible configuration of the fluctuating background field. For this purpose we used the replica trick, and then integrated out the fluctuations. As a result, a new effective interaction between charged currents arises in the averaged Lagrangian density. This interaction is proportional to the magnitude of the auto-correlation function in the electromagnetic fluctuations and to the charge of the field $q^2\Delta$. 

Regarding the self-energy calculations, in the weak electric field limit, the divergent amplitudes remain the same as for the bare $\lambda\phi^4$ theory, which are reduced to the vacuum diagram. This is a consequence of the perturbative treatment of the weak electric field propagator, which in the limit $|qE|\rightarrow0$ recovers the expression for the free scalar propagator in vacuum. The spectral density function in this weak electric field regime shows that the self-interaction increases the value of the pole mass, whereas the fluctuations produce quasi-particle states of finite life-time inversely proportional to $\Delta$. On the other hand, our results for the strong electric field regime show that the self-energy is directly proportional to the electric field in this limit, and free of logarithmic UV divergences. As a consequence, after we add the corresponding counterterms to the Lagrangian, the renormalized self-energy vanishes in the very strong electric field regime. Hence, in the perturbative approach, the physical mass is not modified by the presence of electromagnetic noise.

The vertex correction at one loop allow us to find the vertex counterterms, and consequently, the renormalization group equations for both couplings. The RGE flow shows that both parameters, $\lambda(t)$ and $\tilde{\Delta}(t)$, grow with the logarithmic energy scale $t$ in the weak electric field regime, with the noise working as a damping factor for the self-interaction strength. In contrast, in the very strong electric field regime, the beta functions show the emergence of asymptotic freedom, thus implying that the scalar fields behave as free particles at high energy values under these conditions. The later effect is expected from a simple heuristic argument: for two identical charged particles, a very strong background electric field will apply a local force to separate them, thus weakening the self-interacting effects and leading to an effective a quasi-free particle description. 
Future directions to explore in this model are the finite temperature effects, as well as the interpolation between the weak and very strong electric field regimes by means of the full propagator defined in Eq.~\eqref{eq_pro_full}, which are currently under development by us and will be communicated elsewhere.

\begin{acknowledgments}
M. M. V. acknowledges the financial support of ANID National Ph.D. scholarship No. 21250633. E. M. acknowledges financial support from ANID Fondecyt Grant No. 1230440. M.L. acknowledges the financial support of ANID Fondecyt Grants No. 1260183, No. 1250206, and No. 1241436.
\end{acknowledgments}

%

\appendix
\begin{widetext}
\section{The propagator for a very weak electric field}
\label{Ap_weak_field}
From the exact Schwinger proper time parametrization for the scalar propagator,
\begin{align}
D_E(p) = \int_0^\infty ds\, \frac{e^{-s\left[\frac{\tanh(qEs)}{qEs}p_\parallel^2 + p_\perp^2 + m^2\right]}}{\cosh(qEs)},
\label{Ap_propSchwing}
\end{align}
it is straightforward to derive an explicit power series representation in terns of the electric field $\varepsilon = q E$, as follows
\begin{eqnarray}
D_E(p,\varepsilon) &=& D_E(p,0) + \varepsilon\frac{\partial}{\partial\varepsilon}D_E(p,\varepsilon =0)+ \frac{\varepsilon^2}{2}\frac{\partial^2}{\partial\varepsilon^2}D_E(p,\varepsilon=0) + O(\varepsilon^3).
\end{eqnarray}
From the trivial identities $\lim_{\varepsilon\rightarrow 0}\cosh(s\varepsilon) = 1$, and $\lim_{\varepsilon\rightarrow 0}\tanh(s\varepsilon)/(s\varepsilon) = 1$, we obtain the first term
\begin{eqnarray}
D_E(p,0) &=& \lim_{\varepsilon\rightarrow 0} D_E(p,\varepsilon) \notag\\[5pt] &=& \int_0^\infty ds\, e^{-s\left[p^2 + m^2\right]}\nonumber\\[5pt]
&=& \frac{1}{p^2 + m^2}.
\end{eqnarray}
For the next term, we notice that
\begin{eqnarray}
\lim_{\varepsilon\rightarrow 0}\frac{\partial}{\partial\varepsilon}D_E(p,\varepsilon) = \lim_{\varepsilon\rightarrow 0} \int_0^{\infty}ds\,  e^{-s\left[\frac{\tanh(s\varepsilon)}{s\varepsilon}p_\parallel^2 + p_\perp^2 + m^2\right]} \frac{s}{\cosh(s\varepsilon)}\left[ -\tanh(s\varepsilon) + \frac{p_{\parallel}^2}{\varepsilon}\left( \frac{\tanh(s\varepsilon)}{s\varepsilon} - 1  \right)  \right] = 0.
\end{eqnarray}
So the first-order contribution in the electric field vanishes exactly.
Finally, the second-order term is given by
\begin{eqnarray}
\lim_{\varepsilon\rightarrow 0} \frac{\varepsilon^2}{2}\frac{\partial^2}{\partial\varepsilon^2}D_E(p,\varepsilon) = \int_0^{\infty}ds e^{-s\left[p^2 + m^2\right]} \left( \frac{2}{3} p_{\parallel}^2 s^3 - s^2 \right),\nonumber\\
\end{eqnarray}
and applying the elementary identity, for $n\in\mathbb{N}_0$,
\begin{eqnarray}
\int_0^{\infty}ds e^{-s \alpha} s^n = \frac{\Gamma(n+1)}{\alpha^{n+1}} = \frac{n!}{\alpha^{n+1}},
\end{eqnarray}
we obtain
\begin{eqnarray}
\lim_{\varepsilon\rightarrow 0} \frac{\varepsilon^2}{2}\frac{\partial^2}{\partial\varepsilon^2}D_E(p,\varepsilon) &=& \frac{2}{3}\frac{3! p_{\parallel}^2}{(p^2 + m^2)^4} - \frac{2!}{(p^2+m^2)^3}\nonumber\\[5pt]
&=& 2\left[ \frac{2 p_{\parallel}^2}{(p^2 + m^2)^4} - \frac{1}{(p^2+m^2)^3} \right].\nonumber\\
\end{eqnarray}
Therefore, substituting these partial results into the Taylor expansion in Eq.~\eqref{Ap_propSchwing}, we obtain the final form for the weak field scalar propagator
\begin{eqnarray}
D_E(p) &=& \frac{1}{p^2 + m^2} + (q E)^2 \left[ \frac{2 p_{\parallel}^2}{(p^2 + m^2)^4} - \frac{1}{(p^2+m^2)^3} \right]+ O(qE)^4.
\end{eqnarray}

\section{Integrals involved in the self-energy calculations}
\label{AppSelfEnergy}
Recall that we defined the dimensionless coupling
\begin{align}
\tilde{\Delta} = q^2m^2\Delta.
\end{align}
As discussed in the main text, the total self energy involves two additive contributions proportional to each coupling
\begin{align}
\Sigma(p,E) = -\frac{\lambda}{2} &\int \frac{d^4k}{(2\pi)^4} D_E(k)- 2i\frac{\tilde{\Delta}}{m^2} \int \frac{d^4k}{(2\pi)^4}\,D_E(k)\left(p-k\right)^2.
\end{align}

Since we considered two different limits for the scalar propagator, we shall compute the integrals in both limits separately.

\subsection{Case (A): Very weak field limit $|qE|/m^2\ll 1$}
The self energy contribution arising from the $\lambda\phi^4$ interaction is given by
\begin{align}
\Sigma_\lambda(E) 
&= -\frac{\lambda}{2}\int \frac{d^4k}{(2\pi)^4} \left\{\frac{1}{k^2+m^2} + (qE)^2\left[\frac{-1}{(k^2+m^2)^3} + \frac{2k_\parallel^2}{(k^2+m^2)^4}\right]\right\}.
\end{align}
The first integral is logarithmically divergent, and can be evaluated using dimensional regularization for $\epsilon=4-d \to 0^+$
\begin{align}
\int \frac{d^4k}{(2\pi)^4} \frac{1}{k^2+m^2} &= \mu^\epsilon \frac{d^dk}{(2\pi)^d} \frac{1}{k^2+m^2}\notag \\[5pt]
&= \frac{1}{(4\pi)^2}\Gamma(1-\epsilon/2)\left(\frac{4\pi\mu^2}{m^2}\right)^{\epsilon/2}\notag\\[5pt]
&= -\frac{m^2}{(4\pi)^2}\left[\frac{2}{\epsilon} - \gamma + \ln\left(\frac{4\pi\mu^2}{m^2}\right) + 1 \right],\label{A1-self}
\end{align}
where $\gamma\approx 0.5772$ is the Euler-Mascheroni constant and $\mu$ is the energy scale. The second integral is finite at $d=4$, resulting
\begin{align}
\int \frac{d^4k}{(2\pi)^4} \frac{1}{\left(k^2+m^2\right)^3} &= \mu^\epsilon \frac{d^dk}{(2\pi)^d} \frac{1}{\left(k^2+m^2\right)^3}\notag\\[5pt]
&= \mu^\epsilon \frac{\left(m^2\right)^{d/2-3}}{(4\pi)^{d/2}}\frac{\Gamma(3-d/2)}{\Gamma(3)}\notag\\[5pt]
&= \frac{1}{2(4\pi)^2m^2}.\label{A2-self}
\end{align}
For the remaining integral we separate the integrating measure into parallel and perpendicular components as $d^4k = d^2k_\parallel d^2k_\perp$, resulting

\begin{align}
\int \frac{d^4k}{(2\pi)^4} \frac{2k_\parallel^2}{\left(k^2+m^2\right)^4} &=  \int \frac{d^2k_\perp}{(2\pi)^2} \int \frac{d^2k_\parallel}{(2\pi)^2}\frac{2k_\parallel^2}{\left(k_\parallel^2+k_\perp^2+m^2\right)^4} \notag\\[5pt]
&= \mu^\epsilon\int \frac{d^dk_\perp}{(2\pi)^d} \int \frac{d^dk_\parallel}{(2\pi)^d}\frac{2k_\parallel^2}{\left(k_\parallel^2+k_\perp^2+m^2\right)^4},
\end{align}
where $d=2-\epsilon$ in this case. Evaluating we have
\begin{align}
\mu^\epsilon\int \frac{d^dk_\parallel}{(2\pi)^d}\frac{2k_\parallel^2}{\left(k_\parallel^2+k_\perp^2+m^2\right)^4} &= \frac{2}{4\pi}\frac{\Gamma(2)}{\Gamma(4)} \frac{1}{\left(k_\perp^2+m^2\right)^2},
\end{align}
and integrating over the perpendicular components we have
\begin{align}
\int \frac{d^dk_\perp}{(2\pi)^d} \frac{1}{\left(k_\perp^2 + m^2\right)^2} = \frac{1}{4\pi\, m^2}.\label{A3-self}
\end{align}
Therefore, the $\lambda$ contribution to the self energy is given by
\begin{align}
\Sigma_\lambda(E) &= \frac{\lambda m^2}{2\left(4\pi\right)^2}\left[\frac{2}{\epsilon} - \gamma + \ln\left(\frac{4\pi\mu^2}{m^2}\right) + 1 -\frac{(qE)^2}{6m^4}\right].
\end{align}

For the noise contribution we have
\begin{align}
&\Sigma_\Delta(p,E) = -2i\frac{\tilde{\Delta}}{m^2}\int \frac{d^4k}{(2\pi)^4}\left(p^2-2pk+k^2\right)\left\{\frac{1}{k^2+m^2} + (qE)^2\left[\frac{-1}{(k^2+m^2)^3} + \frac{2k_\parallel^2}{(k^2+m^2)^4}\right]\right\}.
\end{align}
The integrals proportional to the external momentum are the same as the ones previously computed at \eqref{A1-self}, \eqref{A2-self} and \eqref{A3-self}. The integrals over $2pk$ vanish, since they are odd functions over an even integration interval. We are left to evaluate the integrals proportional to the factor $k^2$. Using dimensional regularization, the first integral results
\begin{align}
\int \frac{d^4k}{(2\pi)^4} \frac{k^2}{k^2+m^2} &= \mu^\epsilon \int \frac{d^dk}{(2\pi)^d} \frac{k^2}{k^2+m^2}\notag\\[5pt]
&= \frac{m^4}{(4\pi)^2}\left(2-\frac{\epsilon}{2}\right)\Gamma\left(\epsilon/2-2\right)\left(\frac{4\pi\mu^2}{m^2}\right)^{\epsilon/2}\notag\\[5pt]
&= \frac{m^4}{(4\pi)^2}\left[\frac{2}{\epsilon}-\gamma+\ln\left(\frac{4\pi\mu^2}{m^2}\right)+1\right].\label{A3-self}
\end{align}
In the same fashion, the second integral results
\begin{align}
\int \frac{d^4k}{(2\pi)^4} \frac{k^2}{\left(k^2+m^2\right)^3} &= \mu^\epsilon \int \frac{d^dk}{(2\pi)^d} \frac{k^2}{k^2+m^2}\notag\\[5pt]
&=\frac{1}{2(4\pi)^2}\left(2-\frac{\epsilon}{2}\right)\Gamma\left(\epsilon/2\right)\left(\frac{4\pi\mu^2}{m^2}\right)^{\epsilon/2}\notag \\[5pt]
&= \frac{1}{(4\pi)^2}\left[\frac{2}{\epsilon}-\gamma+\ln\left(\frac{4\pi\mu^2}{m^2}\right)-\frac{1}{2}\right].
\end{align}
Finally, for the last integral we must separate into parallel and perpendicular components for $d=2-\epsilon$, resulting in
\begin{align}
\int \frac{d^4k}{(2\pi)^4}\frac{2k^2k_\parallel^2}{\left(k^2+m^2\right)^4} &= \int \frac{d^2k_\perp}{(2\pi)^2} k^2_\perp \int \frac{d^2k_\parallel}{(2\pi)^2}\frac{k_\parallel^2}{\left(k_\parallel^2+k_\perp^2 + m^2\right)^4} + \int \frac{d^2k_\perp}{(2\pi)^2} \int \frac{d^2k_\parallel}{(2\pi)^2}\frac{k_\parallel^4}{\left(k_\parallel^2+k_\perp^2 + m^2\right)^4}\notag\\[5pt]
&=\frac{1}{6(4\pi)}\int \frac{d^2k_\perp}{(2\pi)^2} \frac{k^2_\perp}{\left(k_\perp^2+m^2\right)^2} + \frac{1}{3(4\pi)} \int \frac{d^dk_\perp}{(2\pi)^2} \frac{1}{k_\perp^2+m^2}\notag\\[5pt]
&= \frac{1}{6(4\pi)^2}\left(1-\frac{\epsilon}{2}\right)\Gamma(\epsilon/2)\left(\frac{4\pi\mu^2}{m^2}\right)^{(\epsilon/2)} + \frac{1}{3(4\pi)^2}\Gamma(\epsilon/2)\left(\frac{4\pi\mu^2}{m^2}\right)^{(\epsilon/2)}\notag\\[5pt]
&=\frac{1}{2(4\pi)^2}\left[\frac{2}{\epsilon}-\gamma+\ln\left(\frac{4\pi\mu^2}{m^2}\right) - \frac{1}{3}\right].
\end{align}

Then, the self energy contribution due to the statistical average over electromagnetic fluctuations is given by

\begin{align}
\Sigma_\Delta(p,E) &= \frac{2i\tilde{\Delta}}{(4\pi)^2}\left\{\left(p^2-m^2\right)\left[\frac{2}{\epsilon} - \gamma + \ln\left(\frac{4\pi\mu^2}{m^2}\right) + 1 \right]+\frac{(qE)^2}{6m^4}\right\}.
\end{align}

\subsection{Case (B): Very strong field limit $|qE|/m^2\gg 1$}
In this case, we compute the self-energy contributions by using the very strong limit form of the propagator, Eq.~\eqref{strong-prop}.
\begin{eqnarray}
\Sigma_{\lambda}(p,E) &=& -\frac{\lambda}{2}\int\frac{d^4 k}{(2\pi)^4} D_E(k)\nonumber\\[5pt]
&=& -\frac{\lambda}{2}2\int\frac{d^2 k_{\parallel}}{(2\pi)^2}e^{-\frac{k_{\parallel}^2}{q E}}\int\frac{d^2 k_{\perp}}{(2\pi)^2}\frac{1}{k_{\perp}^2 + m_E^2}\nonumber\\[5pt]
&=& -\lambda\frac{\pi q E}{(2\pi)^2} \mu^{\epsilon}\int\frac{d^d k_{\perp}}{(2\pi)^d}\frac{1}{k_{\perp}^2 +m_E^2}.
\end{eqnarray}
The first integral over $k_{\perp}$ is trivially convergent, while the second has a logarithmic divergence that is handled by dimensional regularization in $d = 2 - \epsilon$, such that we have
\begin{align}
\Sigma_{\lambda}(p,E) &= - \lambda\frac{ q E}{(4\pi)^2}\left[ \frac{2}{\epsilon} - \gamma - \ln\left( \frac{4\pi\mu^2}{m_E^2} \right)  \right].
\end{align}
Similarly, for the other contribution to the self-energy, we have
\begin{eqnarray}
\Sigma_{\Delta}(p,E) &&= \frac{2 i\tilde{\Delta}}{m^2}\int \frac{d^4k}{(2\pi)^4} D_E(k)(p^2 + 2k\cdot p + k^2)\nonumber\\[5pt]
&&= \frac{2 i\tilde{\Delta}}{m^2} \tilde{\Delta} \left\{
\left[p^2 \int \frac{d^2 k_{\parallel}}{(2\pi)^2}e^{-\frac{k_{\parallel}^2}{q E}} + \int \frac{d^2 k_{\parallel}}{(2\pi)^2}e^{-\frac{k_{\parallel}^2}{q E}}k_{\parallel}^2\right]\mu^{\epsilon}\int \frac{d^d k_{\perp}} {(2\pi)^d} \frac{1}{k_{\perp}^2 +m_E^2}
+ \int \frac{d^2 k_{\parallel}}{(2\pi)^2}e^{-\frac{k_{\parallel}^2}{q E}}\mu^{\epsilon}\int \frac{d^d k_{\perp}} {(2\pi)^d} \frac{k_{\perp}^2}{k_{\perp}^2 + m_E^2}
\right\}\nonumber\\[5pt]
&&= \frac{2 i\tilde{\Delta}}{m^2} \left\{
\frac{\left( p^2 (q E) + (q E)^2 \right)}{4\pi}\mu^{\epsilon}\int \frac{d^d k_{\perp}} {(2\pi)^d} \frac{1}{k_{\perp}^2 + m_E^2}
+ \frac{q E}{4\pi} \mu^{\epsilon}\int \frac{d^d k_{\perp}} {(2\pi)^d} \frac{k_{\perp}^2}{k_{\perp}^2 + m_E^2}
\right\}.
\end{eqnarray}

Now, we use the results from Eq. \eqref{A1-self} and \eqref{A3-self} we obtain
\begin{eqnarray}
\Sigma_{\Delta}(p,E) &=& \frac{4 i \tilde{\Delta}}{\left(  4\pi\right)^2} (p^2 - m^2)\frac{(q E)}{m^2}\left[\frac{2}{\epsilon} - \gamma + \ln\left( \frac{4\pi \mu^2}{m_E^2}\right)\right].
\end{eqnarray}

Therefore, the total self-energy in the very strong field limit is given by
\begin{align}
\Sigma(p,E) &= \Sigma_{\lambda}(p,E) + \Sigma_{\Delta}(p,E) \\[5pt] 
&= \frac{1}{(4\pi)^2}\frac{(q E)}{m^2}\left[  
-\lambda m^2 + 4 i \tilde{\Delta}\left(p^2 - m^2\right)
\right]\left[ \frac{2}{\epsilon} - \gamma + \ln\left( \frac{4\pi\mu^2}{m_E^2} \right)  \right].
\end{align}

\section{Renormalized Lagrangian density}\label{renorm-lag}
As shown in the previous Appendix, the integrals involved in the self energy calculations are logarithmically divergent. Hence, we must introduce counterterms in the effective Lagrangian, which can be rewritten in terms of the bare parameters as
\begin{align}
\mathcal{L}_\text{eff} = \left|D_\mu \phi_0^a\right|^2 - m_0^2\left(\phi^{a}_{0}\right)^\dagger\phi_0^a - \frac{1}{4}\lambda_0\left[\left(\phi^{a}_{0}\right)^\dagger\phi_0^a\right]^2- \frac{i\tilde{\Delta}}{m^2}\,j_{\mu0}^a j_{b0}^\mu,
\end{align}
where the sum over replica indices $a$ and $b$ is implicit and the currents depend on the bare field as
\begin{align}
j_{\mu}(x) = \phi_{a0}^\dagger(x)D_\mu\phi_0^a(x) - \left(D_\mu\phi_0^a(x)\right)^\dagger\phi_0^a(x).
\end{align}
Rescaling the bare field as
\begin{align}
\phi_a^{(0)}(x) = \sqrt{Z} \phi_a(x),
\end{align}
the Lagrangian density results
\begin{align}
\mathcal{L}_\text{eff} = Z\left|D^\mu\phi^a\right|^2 - m_0^2 Z\phi^\dagger_a\phi^a - \frac{\lambda_0}{4}Z^2\left(\phi^\dagger_a\phi_2\right)^2- \frac{i\tilde{\Delta_0}}{m^2}j_\mu^aj^\mu_b.
\end{align}
Then, if we define the following counterterms for the mass, field strength and couplings as
\begin{align}
&\delta_Z = Z-1\,,\label{delta-z}\\
&\delta_{m^2}= m_0^2 Z - m^2\,,\\
&\delta_\lambda = \lambda_0 Z^2 - \lambda,\label{delta-lambda}\\
&\delta_{\tilde{\Delta}} = {\tilde{\Delta}}_0 Z^2 - {\tilde{\Delta}} \label{delta-delta},
\end{align}
and the full renormalized Lagrangian will be given by
\begin{align}
\mathcal{L}_\text{eff} =& \sum_{a=1}^n \left[\big|D_\mu \phi_a\big|^2 -  m(\phi_a)^\dagger\phi_a - \frac{1}{4}\lambda\left(\phi_a^\dagger\phi_a\right)^2 \right] - i\tilde{\Delta}\sum_{a,b}^n j_{\mu,a} j^{\mu,b}\notag\\[5pt]
& + \sum_{a=1}^n \left[\delta_Z\big|D_\mu \phi_a\big|^2 - \delta_{m^2} \phi_a^\dagger\phi_a - \frac{1}{4}\delta_\lambda\left(\phi_a\phi_a\right)^2 \right] + i\delta_{\tilde{\Delta}}\sum_{a,b}^n j_{\mu,a} j^{\mu,b} \label{nuevo_l}.
\end{align}

\section{Integrals involved in the vertex corrections}
\label{AppVertex}

\subsection{Case (A): Very weak field limit $|qE|/m^2\ll 1$}
As discussed in the main text, the $s-$channel contribution to the vertex function is given by the integral
\begin{align}
iV(p,E) = {\int \frac{d^4k}{(2\pi)^4}}\,iD_E(k)\,iD_E(p+k).
\end{align}
Using the propagator for the weak field limit, given by Eq. \eqref{weak-prop}, the vertex function results
\begin{align}
iV(p,E) = \int& \frac{d^4k}{(2\pi)^4}\left\{\frac{1}{k^2+m^2} + (qE)^2\left[\frac{-1}{(k^2+m^2)^3} + \frac{2k_\parallel^2}{(k^2+m^2)^4}\right]\right\}\notag\\[5pt]
&\times \left\{\frac{1}{(p+k)^2+m^2} + (qE)^2\left[\frac{-1}{\left[(p+k)^2+m^2\right]^3} + \frac{2(p+k)_\parallel^2}{\left[(p+k)^2+m^2\right]^4}\right]\right\}.
\end{align}
As we want to find the beta functions for this theory, we only need to compute the divergent contribution, which is given by the integral
\begin{align}
A_1 = \int\frac{d^4k}{(2\pi)^4} \frac{1}{\left[(p+k)^2+m^2\right](k^2+m^2)}.
\end{align}

Using Feynman parametrization and dimensional regularization, we have
\begin{align}
A_1 = \int_0^\infty dx\int\frac{d^d l}{(2\pi)^d} \frac{1}{\left(l^2+\kappa\right)^2},
\end{align}
where
\begin{align}
l &= k + xp,\\[5pt]
\kappa &= xp^2(1-x)+m^2.
\end{align}
Using dimensional regularization with $\epsilon = 4 - d\rightarrow 0^{+}$, we find the following
\begin{align}
A_1 &= \frac{1}{(4\pi)^2}\left[\frac{2}{\epsilon}-\gamma+\ln\left(\mu^2\right)\right]+ \frac{1}{(4\pi)^2}\int_0^1dx \ln\left(\frac{4\pi}{xp^2(1-x)+m^2}\right).
\end{align}

Therefore, the divergent vertex function is
\begin{align}
iV(p^2,0) = \frac{4ip^2}{(4\pi)^2}\left[\frac{2}{\epsilon}-\gamma+\ln\left(\mu^2\right)\right]+\text{finite terms}.
\end{align}
The vertex function for both vertex result
\begin{align}
iV_{\lambda}(p,E) &= -\frac{i}{2}\cdot iV(p,E),\\[5pt]
iV_{\Delta}(p,E) &= -4ip^2 \cdot iV(p,E)
\end{align}
The scattering amplitude involving the renormalized vertex is defined as
\begin{align}
i\mathcal{M}_g(p_1p_2\to p_3p_4) = 
- ig + (-ig)^2\left[iV_g(4m^2,E)+2iV_g(0,E)\right]+i\delta_g.
\end{align}
Hence, the counterterms for both couplings result
\begin{align}
\delta_\lambda &= -\frac{3\lambda^2}{2(4\pi)^2}\left[\frac{2}{\epsilon}-\gamma+\ln\left(\mu^2\right)\right],\\[5pt]
\delta_\Delta &= -\frac{\tilde{\Delta}^2}{\pi^2}\left[\frac{2}{\epsilon}-\gamma+\ln\left(\mu^2\right)\right] .
\end{align}

\subsection{Case (B): Very strong field limit $|qE|/m^2 \gg 1$}
In the strong field case we use the propagator from Eq. \eqref{strong-prop}. Then the vertex function results
\begin{align}
iV(p,E) &= \int \frac{d^4k} {(2\pi)^4} \frac{2e^{-\frac{k_{\parallel}^2}{q E}}}{k_{\perp}^2 + q E + m^2} \frac{2e^{-\frac{\left(p_{\parallel}+k_{\parallel}\right)^2}{q E}}}{\left(p_\perp + k_{\perp}\right)^2 + q E + m^2} .
\end{align}
Separating the integral in the perpendicular and parallel components we get
\begin{align}
iV(p,E) &= 4\int \frac{d^2k_\parallel} {(2\pi)^2}  e^{\frac{-k_{\parallel}^2-\left(p_{\parallel}+k_{\parallel}\right)^2}{q E}}\int \frac{d^2k_\perp} {(2\pi)^2} \frac{1}{k_{\perp}^2 + m_E^2}\cdot\frac{1}{\left(p_\perp + k_{\perp}\right)^2 +m_E^2}.
\end{align}
The first integral is quadratic in $k_\parallel$, and can be solved by completing squares as
\begin{align}
-k_{\parallel}^2-\left(p_{\parallel}+k_{\parallel}\right)^2&= 2\left(k_\parallel + \frac{p_\parallel^2}{2}\right)^2 + \frac{p_\parallel}{2} \notag\\
&= 2l_\parallel^2 + \frac{p_\parallel^2}{2},
\end{align}
where $l_\parallel = k_\parallel + \frac{p_\parallel}{2}$. Then, the integral over the parallel component results
\begin{align}
A_1 &= \int \frac{d^2k_\parallel} {(2\pi)^2}  e^{\frac{-k_{\parallel}^2-\left(p_{\parallel}+k_{\parallel}\right)^2}{q E}}\\[5pt] 
&= e^{-\frac{p_\parallel^2}{2qE}} \int \frac{d^2k_\parallel} {(2\pi)^2}  e^{-\frac{2 l_\parallel^2}{q E}} \notag\\[5pt]
&= \frac{e^{-\frac{p_\parallel^2}{2qE}}}{(2\pi)^2} \frac{qE\pi}{2}.
\end{align}

The integral over the perpendicular components is given by
\begin{align}
A_2 = \int \frac{d^2k_\perp} {(2\pi)^2} \frac{1}{k_{\perp}^2 + m_E^2}\cdot\frac{1}{\left(p_\perp + k_{\perp}\right)^2 +m_E^2},
\end{align}
which can be solved with Feynman parameters as
\begin{align}
A_2 = \int_0^1 dx \int \frac{d^2l_\perp} {(2\pi)^2} \frac{1}{\left(l_\perp^2 + \kappa\right)^2},
\end{align}
where
\begin{align}
l_\perp &= k_\perp + xp_\perp, \\
\kappa &= x(1-x)p_\perp^2 + m_E^2.
\end{align}
This integral is regular in two dimensions, resulting in
\begin{align}
A_2 = \frac{1}{4\pi}\int_0^1 dx \frac{1}{x(1-x)p_\perp^2 + m_E^2}.
\end{align}
To solve this integral we factorize the denominator as
\begin{align}
x(1-x)p_\perp^2 + m_E^2 = -p_\perp^2(x-x_1)(x-x_2),
\end{align}
where
\begin{align}
x_{1,2}&= \frac{1}{2} \pm \sqrt{1+\frac{4m_E^2}{p_\perp^2}}.
\end{align}
Then, by partial fraction decomposition we get
\begin{align}
A_2 &= -\frac{1}{4\pi p_\perp^2} \int_0^1 dx \frac{1}{(x-x_1)(x-x_2)} \notag\\[5pt]
&=-\frac{1}{4\pi p_\perp^2(x_2-x_1)} \int_0^1 dx \left[\frac{1}{(x-x_2)} - \frac{1}{(x-x_1)}\right]\notag\\[5pt]
&=-\frac{1}{4\pi p_\perp^2(x_2-x_1)} \ln\left(\frac{1-x_2}{1-x_1}\cdot\frac{x_1}{x_2}\right).
\end{align}
Finally, the vertex function results
\begin{align}
iV(p,E) = \frac{qE}{2(4\pi p_\perp)^2}\frac{e^{-\frac{p_\parallel^2}{2qE}}}{(x_2-x_1)}\ln\left(\frac{1-x_2}{1-x_1}\cdot\frac{x_1}{x_2}\right).
\end{align}

\section{Solution of the running couplings}
\label{Ap_running_couplings}
\subsection{Weak electric field regime}

In this section, we consider the analytical solution to the system of differential equations for the couplings $g = (\lambda,\tilde{\Delta})$, 
\begin{align}
&\beta_{\tilde{\Delta}} = \frac{2}{\pi^2}\tilde{\Delta}^2\left(1-\frac{i}{4}\right)\label{apbd},\\
&\beta_\lambda = \frac{3}{16\pi^2}\lambda^2 - \frac{i}{2\pi^2}\lambda\tilde{\Delta}.\label{apbl}
\end{align}
We first notice that the first two equations, Eq.~\eqref{apbd} and Eq.~\eqref{apbl}, constitute an autonomous subsystem, that we solve first. We start by solving Eq.~\ref{apbd} by separation of variables, as follows
\begin{eqnarray}
\int_{\tilde{\Delta}_0}^{\tilde{\Delta}(t)} \frac{d\tilde{\Delta}}{\tilde{\Delta}^2} = \frac{2}{\pi^2}\left( 1 - \frac{i}{4} \right)\int_0^t dt'
\end{eqnarray}
such that we obtain
\begin{eqnarray}
\frac{1}{\tilde{\Delta}_0} - \frac{1}{\tilde{\Delta}(t)} = \frac{2}{\pi^2}\left( 1 - \frac{i}{4} \right) t
\end{eqnarray}
and after some elementary algebra, we solve in favor of $\tilde{\Delta}(t )$,
\begin{eqnarray}
\tilde{\Delta}(t) = \frac{\tilde{\Delta}_0}{1 - \frac{2}{\pi^2}\left( 1 - \frac{i}{4} \right)\tilde{\Delta}_0 t} 
\label{eq_apDelta}
\end{eqnarray}
Let us now consider Eq.~\eqref{apbl}, dividing by $\lambda^2$ and reorganizing the terms as follows
\begin{eqnarray}
-\frac{1}{\lambda(t)^2}\frac{d\lambda}{dt} - \frac{i}{2\pi^2}\frac{\tilde{\Delta}(t)}{\lambda(t)} = -\frac{3}{16\pi^2}
\label{eq_apaux}
\end{eqnarray}
Let us now introduce the auxiliary function
\begin{equation}
u(t) \equiv \frac{1}{\lambda(t)},
\end{equation}
so that the differential Eq.~\eqref{eq_apaux} becomes
\begin{equation}
\frac{du}{dt} - \frac{i}{2\pi^2}\tilde{\Delta}(t)u(t) = -\frac{3}{16\pi^2}
\end{equation}
An integrating factor for this equation is the function
\begin{equation}
\eta(t) = \exp\left[ -\frac{i}{2\pi^2}\int_0^t dt' \tilde{\Delta}(t')\right]
\label{eq_apeta1}
\end{equation}
such that
\begin{eqnarray}
\frac{d}{dt}\left[ u(t)\eta(t) \right] = -\frac{3}{16\pi^2}\eta(t).
\end{eqnarray}
The exact solution then is given after direct integration, as follows
\begin{eqnarray}
\int_0^t \frac{d}{dt'}\left[ u(t')\eta(t') \right] dt' &=& u(t)\eta(t) - u(0)\nonumber\\
&=&  -\frac{3}{16\pi^2}\int_0^t \eta(t')dt',
\end{eqnarray}
where we used the obvious condition $\eta(t=0) = 1$.
Solving for $u(t) = 1/\lambda(t)$, after some elementary algebra, we obtain
\begin{eqnarray}
\frac{1}{\lambda(t)} = \frac{1}{\eta(t)} \left[\frac{1}{\lambda_0} - \frac{3}{16\pi^2}\int_0^t \eta(t')dt'\right].
\end{eqnarray}
After some elementary algebra, we solve explicitly for $\lambda(t)$
\begin{eqnarray}
\lambda(t) = \frac{\lambda_0 \eta(t)}{1 - \frac{3}{16\pi^2}\lambda_0\int_0^t \eta(t')dt'}.
\label{eq_aplambda1}
\end{eqnarray}
All the effects of the electric noise over the effective coupling $\lambda(t)$ are contained in the function $\eta(t)$. Now, we solve explicitly $\eta(t)$, using the exact solution Eq.~\eqref{eq_apDelta}, to obtain
\begin{eqnarray}
\eta(t) &=& \exp\left[ -\frac{i\tilde{\Delta}_0}{2 \pi^2}\int_0^t\frac{dt'}{1 - \frac{2}{\pi^2}\left( 1 - \frac{i}{4} \right)\tilde{\Delta}_0 t'}\right]\nonumber\\[5pt]
&=&  \exp\left[\frac{i}{4 - i}\ln\left( 1 - \frac{2}{\pi^2}\left( 1 - \frac{i}{4} \right)\tilde{\Delta}_0 t \right)\right]\nonumber\\[5pt]
&=& \left[ 1 - \frac{2}{\pi^2}\left( 1 - \frac{i}{4} \right)\tilde{\Delta}_0 t \right]^{\frac{i}{4 - i}}.
\label{eq_apeta2}
\end{eqnarray}
Similarly, we trivially obtain the integral
\begin{eqnarray}
\int_0^t \eta(t')dt' &=& \int_0^t \left( 1 - \frac{2}{\pi^2}\left( 1 - \frac{i}{4} \right)\tilde{\Delta}_0 t' \right)^{\frac{i}{4 - i}} dt'\nonumber\\[5pt]
&=& \frac{\pi^2}{2\tilde{\Delta}_0}\left[1 - \left( 1 - \frac{2}{\pi^2}\left( 1 - \frac{i}{4} \right)\tilde{\Delta}_0 t \right)^{\frac{4}{4 - i}}\right]
\label{ap_inteta}
\end{eqnarray}
We can finally solve explicitly for the running coupling $\lambda(t = 1/2\ln(p^2/m^2))$, by inserting Eq.~\eqref{eq_apeta2} and Eq.~\eqref{ap_inteta} into Eq.~\eqref{eq_aplambda1}
\bea
\lambda(t) = \frac{\lambda_0  \left( 1 - \frac{2}{\pi^2}\left( 1 - \frac{i}{4} \right)\tilde{\Delta}_0 t \right)^{\frac{i}{4 - i}} }{1 - \frac{3\lambda_0 }{16\pi^2}\frac{\pi^2}{2\tilde{\Delta}_0}\left[1 - \left( 1 - \frac{2}{\pi^2}\left( 1 - \frac{i}{4} \right)\tilde{\Delta}_0 t \right)^{\frac{4}{4 - i}}\right]  }\nonumber\\
\eea

\subsection{Very strong electric field regime}\label{strong-running}
The beta functions for the couplings and the physical mass parameters are given by
\begin{align}
\beta_{\tilde{\Delta}} &= -i\frac{qE}{\pi^2m^2}\tilde{\Delta}^2,\\[5pt]
\beta_\lambda &= -i\frac{qE}{\pi^2m^2} \lambda \tilde{\Delta}
\end{align}
The first two equations can be solved by direct integration. The solution to the $\tilde{\Delta}$ equation can be obtained by separation of variables as
\begin{align}
\int_{\tilde{\Delta}_0}^{\tilde{\Delta}(t)}\frac{d\tilde{\Delta}}{\left(\tilde{\Delta}\right)^2} = -i\frac{qE}{\pi^2m^2} \int_0^t dt',
\end{align}
resulting in
\begin{align}
\frac{1}{\tilde{\Delta}(t)} - \frac{1}{\tilde{\Delta}_0} = i\frac{qE}{\pi^2m^2}t.
\end{align}
Then, the running coupling $\tilde{\Delta}(t)$ results
\begin{align}
\tilde{\Delta}(t) = \frac{\tilde{\Delta}_0}{1 + i\frac{qE}{\pi^2m^2}\tilde{\Delta}_0 t}.
\end{align}

With this solution, the equation for $\lambda$ results
\begin{align}
\frac{d\lambda}{dt} = -i\frac{qE}{\pi^2m^2} \lambda(t)\cdot \frac{\tilde{\Delta}_0}{1 + i\frac{qE}{\pi^2m^2}\tilde{\Delta}_0 t}.
\end{align}

This equation can be solved by separation of variables, resulting in
\begin{align}
\int_{\lambda_0}^{\lambda(t)} \frac{dt}{\lambda} = -i\frac{qE}{\pi^2m^2}\tilde{\Delta}_0 \int_0^t \frac{dt'}{1 + i\frac{qE}{\pi^2m^2}\tilde{\Delta}_0 t}.
\end{align}
Direct integration yields
\begin{align}
\ln\left(\frac{\lambda(t)}{\lambda_0}\right) = \ln\left(\frac{1}{1+i\frac{qE}{\pi^2m^2}\tilde{\Delta}_0t}\right).
\end{align}
Finally, the coupling $\lambda(t)$ evolves with the energy scale $t$ as
\begin{align}
\lambda(t) = \frac{\lambda_0}{\tilde{\Delta}_0}\tilde{\Delta}(t).
\end{align}
\end{widetext}
\end{document}